\documentstyle[aps,prb,epsfig,floats]{revtex}
\begin{document}
\draft \wideabs{
\title{Stability of Trions in Strongly Spin Polarized Two-Dimensional
Electron Gases }
\author{S. A. Crooker}
\address{National High Magnetic Field Laboratory - LANL, MS E536, Los Alamos, NM 87545}
\author{E. Johnston-Halperin and D. D. Awschalom}
\address{Department of Physics, University of California,
Santa Barbara CA 93106}
\author{R. Knobel and N. Samarth}
\address{Department of Physics, Pennsylvania State University,
 University Park PA 16802}
\date{Submitted 5 January 2000}
\maketitle
\begin{abstract}
Low-temperature magneto-photoluminescence studies of negatively
charged excitons (${X^-_s}$ trions) are reported for {\it n}-type
modulation-doped ZnSe/Zn(Cd,Mn)Se quantum wells over a wide range
of Fermi energy and spin-splitting. The magnetic composition is
chosen such that these magnetic two-dimensional electron gases
(2DEGs) are highly spin-polarized even at low magnetic fields,
throughout the entire range of electron densities studied
(${5\times10^{10}}$ to ${6.5\times10^{11}}$ cm${^{-2}}$). This
spin polarization has a pronounced effect on the formation and
energy of ${X^-_s}$, with the striking result that the trion
ionization energy (the energy separating ${X^-_s}$ from the
neutral exciton) follows the temperature- and magnetic
field-tunable Fermi energy. The large Zeeman energy destabilizes
${X^-_s}$ at the ${\nu=1}$ quantum limit, beyond which a new PL
peak appears and persists to 60 Tesla, suggesting the formation of
spin-triplet charged excitons.
\end{abstract}
} \narrowtext Magnetic two-dimensional electron gases (2DEGs)
represent a relatively new class of semiconductor quantum
structure in which an electron gas is made to interact strongly
with embedded magnetic
moments.\cite{Smorchkova,Salib,Haury,Wojtowicz} Typically,
magnetic 2DEGs (and 2D hole gases) are realized in
modulation-doped II-VI diluted magnetic semiconductor quantum
wells in which paramagnetic spins (Mn${^{2+}}$, ${S=\frac{5}{2}}$)
interact with the confined electrons via a strong ${J_{s-d}}$
exchange interaction.\cite{Awschalom} This interaction leads to an
enhanced spin splitting of the electron Landau levels which
follows the Brillouin-like Mn${^{2+}}$ magnetization, saturating
in the range 10-20 meV by a few Tesla. Since the spin splitting
can greatly exceed both the cyclotron (${\simeq}$1 meV/T) and
Fermi energies, these magnetic 2DEGs consist largely of
spin-polarized Landau levels, and serve as interesting templates
for studies of quantum transport in the absence of spin
gaps.\cite{Smorchkova} In addition, it has been recognized that
this interplay between the cyclotron, Zeeman and Fermi energies
may also be exploited in magneto-optical experiments to gain
insights into the rich spectrum of optical excitations found in
2DEGs.\cite{Wojtowicz} The aim of this paper is to use strongly
spin-polarized magnetic 2DEGs, containing a wide range of electron
densities, to shed new light on the spin-dependent properties of
negatively charged excitons (or trions).

Predicted in 1958 by Lampert\cite{Lampert} and first observed by
Kheng\cite{Kheng} in 1993, the singlet state of the negatively
charged exciton (the ${X^-_s}$ trion) consists of a spin-up and
spin-down electron bound to a single hole.\cite{Wojtowicz} The
energy required to remove one of these electrons (leaving behind a
neutral exciton ${X^0}$) is the ${X^-_s}$ ionization energy
${\Delta E_X}$, usually defined as the energy between ${X^-_s}$
and ${X^0}$ features in optical studies. ${\Delta E_X}$ is small;
typically only ${\sim}$1 meV, ${\sim}$3 meV, and ${\sim}$6 meV in
GaAs-\cite{Shields}, CdTe-\cite{Kheng}, and
ZnSe-based\cite{Astakhov} 2DEGs respectively. The spin-singlet
nature of the two electrons in ${X^-_s}$ suggests that ${\Delta
E_X}$ -- and hence trion stability -- should be sensitive to the
Zeeman energy and spin-polarization of the 2DEG. Here, we
explicitly study highly spin-polarized magnetic 2DEGs to establish
empirical correlations between Zeeman energy and trion stability
over a broad range of carrier densities. In particular,
magneto-photoluminescence (PL) measurements demonstrate the
striking result that ${\Delta E_X}$ follows the energy of the
Fermi surface, which can be tuned independently from the Landau
levels via the strong Zeeman dependence on temperature and applied
field. The role of the Fermi and Zeeman energies in determining
${\Delta E_X}$ is studied for all carrier densities, and
qualitative agreement with numerical calculations is found. The
giant spin-splitting in these systems is found to reduce ${\Delta
E_X}$, eventually driving a rapid suppression of ${X^-_s}$ by the
${\nu=1}$ quantum limit, beyond which the formation of a new peak
in the PL (which persists to 60T) may signify the formation of
spin-triplet charged excitons.

These experiments are performed at the National High Magnetic
Field Laboratory, in the generator-driven 60 Tesla Long-Pulse
magnet and a 40T capacitor-driven magnet (with 2000 ms and 500 ms
pulse duration, respectively), as well as a 20T superconducting
magnet. Light is coupled to and from the samples via single
optical fibers (200${\mu m}$ or 600${\mu m}$ diameter), and
excitation power is kept below 200${\mu W}$. Thin-film circular
polarizers between the fiber and sample permit
polarization-sensitive PL studies.  In the pulsed magnet
experiments, a high-speed CCD camera acquires complete optical
spectra every 1.5 ms, enabling reconstruction of the entire
spectra vs. field dependence in a single magnet
shot.\cite{Crooker1} The magnetic 2DEG samples are MBE-grown {\it
n}-type modulation-doped 105${\AA}$ wide single quantum wells into
which Mn${^{2+}}$ are ``digitally" introduced in the form of
equally-spaced fractional monolayers of MnSe.  Specifically, the
quantum wells are paramagnetic digital alloys of
(Zn${_{1-x}}$Cd${_x}$Se)${_{m-f}}$(MnSe)${_f}$ with {\it x}= 0.1
to 0.2, m=5 and f=1/8 or 1/16 effective monolayer
thickness.\cite{Smorchkova} The electron densities, determined
from Shubnikov-deHaas (SdH) oscillations in transport, range
between ${5\times10^{10}}$ and ${6.5\times10^{11}}$ cm${^{-2}}$.
All samples show a large spin splitting at 1.5 K, with
``effective" g-factors in the range
${70<g_e^{eff}(H\rightarrow0)<100}$.
\begin{figure}
\epsfxsize=2.5in \center \epsfbox{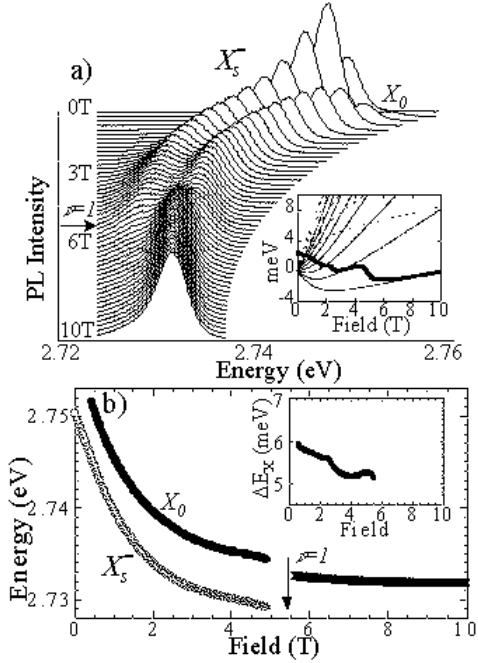} \vspace{0.1in}
\caption{a) Characteristic evolution of the PL spectra at 1.5K in
low-density (${n_e=1.24\times10^{11}}$cm${^{-2}}$) magnetic 2DEGs,
showing a collapse of the ${X^-_s}$ and ${X^0}$ peaks at
${\nu=1}$. Inset: spin-up (dotted) and spin-down (solid) LLs, and
Fermi energy in this sample. b) PL peak energies. Inset: the
${X^-_s}$-${X^0}$ energy splitting, which follows the Fermi
energy.} \label{fig1}
\end{figure}

Figure 1a shows the evolution of the PL spectra in a magnetic 2DEG
with relatively low carrier density ${1.24\times10^{11}}$
cm${^{-2}}$ and ${g_{eff}=73}$ at 1.5K. This sample has a mobility
of 14000 cm${^2}$/Vs and exhibits clear SdH oscillations in
transport.\cite{Knobel} At ${H=0}$, the data show a strong PL peak
at 2.74 eV with a small satellite ${\sim}$6 meV higher in energy.
With applied field, the peaks shift rapidly to lower energy in the
${\sigma ^+}$ polarization due to the large Zeeman energy (the
${\sigma ^-}$ emission disappears completely at low fields in all
the magnetic 2DEGs, much like their undoped
counterparts\cite{Crooker2}). By 1 T, the satellite develops into
a clear peak of comparable amplitude, and as will be verified in
Fig. 2, we assign the high- and low-energy PL features to ${X^0}$
and ${X^-_s}$. At ${\nu=1}$ (5.5 T), the smooth evolution of the
PL spectra changes abruptly as the ${X^-_s}$ resonance collapses
and a strong, single PL peak emerges at an energy between that of
${X^0}$ and ${X^-_s}$, as shown. This new PL feature persists to
60 T. Fig. 1b shows the energies of the PL peaks (the data are fit
to Gaussians), where the discontinuity at ${\nu=1}$ is clearly
seen. The ${X^-_s}$ ionization energy ${\Delta E_X}$ decreases and
oscillates with magnetic field (inset, Fig 1b). Anticipating Figs.
3 and 4, we note that ${\Delta E_X}$ qualitatively mimics the
Fermi energy in this low-density magnetic 2DEG (plotted in Fig.
1a, inset).

Owing to the giant spin splitting in this sample, the ``ordinary"
Landau level (LL) fan diagram for non-magnetic 2DEGs (with Landau
levels evenly spaced by ${\hbar\omega_c}$, and spin splitting
${\ll\hbar\omega_c}$) is replaced by that shown in the inset of
Fig. 1a. The LLs are simply calculated as
\begin{equation}
\varepsilon_{l,s}=\hbar\omega_c(l+\frac{1}{2})+sE_ZB_{5/2}(5g_{Mn}\mu_BH/2k_BT^*)
\end{equation}
where ${l}$ is the orbital angular momentum index and ${s}$ is the
electron spin (${\pm\frac{1}{2}}$). Here, ${\hbar\omega_c}$ =0.83
meV/T is the electron cyclotron energy, and the second term is the
Zeeman energy: ${B_{5/2}}$ is the Brillouin function describing
the magnetization of the ${S=\frac{5}{2}}$ Mn${^{2+}}$ moments,
${E_Z}$ is the saturation value of the electron splitting,
${g_{Mn}}$=2.0, and ${T^*}$ is an empirical ``effective
temperature" which best fits the low-field energy
shifts.\cite{Awschalom} We ignore the much smaller contribution to
the Zeeman energy arising from the bare electron g-factor.  At low
fields, the spin-down LLs (solid lines) are Zeeman-shifted well
below the spin-up LLs (dotted lines), leading to a highly
spin-polarized electron gas - {\it e.g.}, by 1T, over 95\% of the
electrons are oriented spin-down in this sample. The Fermi energy
${\varepsilon_F}$ (thick line) is calculated numerically by
inverting the integral
\begin{equation} N_e=\int^\infty_{-\infty}
g[\varepsilon,B,T]f[\varepsilon,\varepsilon_F,T]d\varepsilon.
\end{equation}
Here, ${N_e}$ is the known electron density,
${f[\varepsilon,\varepsilon_F,T]}$ is the Fermi-Dirac distribution
and ${g[\varepsilon,B,T]}$ is the density of states, taken to be
the sum of Lorentzian LLs\cite{Potts} of width
${\Gamma=\hbar/2\tau_s}$ centered at the energies
${\varepsilon_{l,s}}$ given in Eq.1. The electron scattering time
${\tau_s}$ is obtained from analyzing SdH oscillations, or
alternatively from the measured mobility.

Typically, identification of ${X^0}$ and ${X^-_s}$ relies on their
polarization properties in reflection or
absorption\cite{Wojtowicz,Kheng}- measurements which directly
probe the available density of states. However in these magnetic
2DEGs, the huge Zeeman splitting and the relatively broad spectral
linewidths (resulting from the high Mn${^{2+}}$ concentration)
complicate these standard analyses. While reflectivity studies in
these samples {\it do} confirm the presence of two bound states at
zero field (as expected for ${X^0}$ and ${X^-_s}$), we rely on
spin-polarized PL excitation measurements to verify the peaks in
finite field, shown in Fig. 2. At fixed field and temperature, we
record the PL while tuning the energy and helicity of the
excitation laser (a frequency-doubled cw Ti:Sapphire laser). Since
the PL is entirely ${\sigma ^+}$ polarized, it must arise from the
recombination of a spin-down (${m_s=-\frac{1}{2}}$) electron with
a ${m_j=-\frac{3}{2}}$ valence hole (see diagram, Fig. 2c). If
that ${m_s=-\frac{1}{2}}$ electron is part of an ${X^-_s}$ trion,
emission will occur at the ${X^-_s}$ energy. Thus the probability
of forming ${X^-_s}$ is related to the number of spin-{\it up}
(${m_s=+\frac{1}{2}}$) electrons present in the system. By
specifically injecting spin-up electrons at the ${\sigma ^-}$
resonance, we do indeed observe an enhancement of the ${X^-_s}$
intensity (Fig. 2a). In contrast, injecting spin-down electrons
with ${\sigma ^+}$ light can (and does) only favor the ${X^0}$
intensity (Fig. 2b). The amplitude ratio, I(${X^-_s}$)/I(${X^0}$),
is plotted in Fig. 2c, where the effects of pumping spin-up and
spin-down electrons are more easily seen. Of related interest, no
difference in this ratio is observed when exciting above the ZnSe
barriers (2.8 eV) - evidence that the injected spin is scrambled
when the electrons spill into the well from the barrier regions.
\begin{figure}
\epsfxsize=2.5in \center \epsfbox{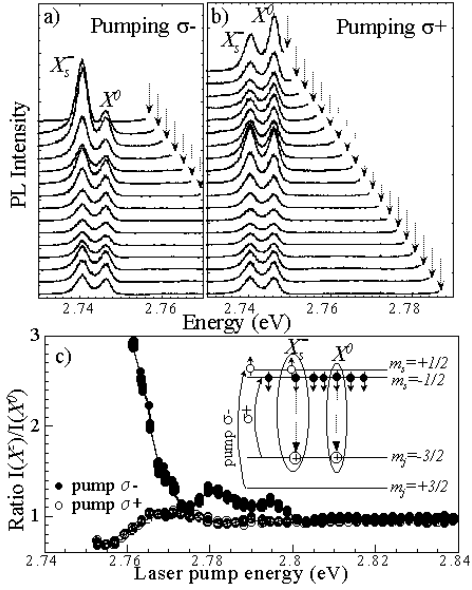} \vspace{0.1in}
\caption{a) PL-excitation at 2.2K and 1 T, showing an enhancement
of ${X^-_s}$ when injecting spin-up electrons on the ${\sigma ^-}$
resonance.  b) A similar enhancement of the ${X^0}$ peak when
injecting spin-down electrons. c) The intensity ratio
I(${X^-_s}$)/I(${X^0}$), with a schematic of the energy levels and
processes involved (the light holes are split off due to quantum
confinement effects).} \label{fig2}
\end{figure}

With the aid of the diagram in Fig. 2c, the evolution of the PL
spectra in Fig. 1 may be interpreted as follows: ${X^-_s}$ and
${X^0}$ are competing channels for exciton formation, with
${X^-_s}$ dominating at zero field. With small applied field, the
large spin-splitting drives a rapid depopulation of the spin-up
electron bands, reducing the probability of ${X^-_s}$ formation
and thus increasing ${X^0}$ formation, as observed. With
increasing field and Zeeman energy, ${X^-_s}$ continues to form
until it is no longer energetically favorable to bind a spin-up
electron -- in this case, evidently, at ${\nu=1}$ when the Fermi
energy falls to the lowest LL.  The PL peak which forms at
${\nu=1}$ (and persists to 60T), with an energy {\it between} that
of ${X^-_s}$ and ${X^0}$, represents formation of a stable new
ground state. A likely candidate is the spin-triplet state of the
negatively charged exciton (${X^-_t}$), wherein both bound
electrons are oriented spin-down. The ${X^-_t}$ trion, predicted
to become the ground state in nonmagnetic 2DEGs at sufficiently
high magnetic field\cite{Shields2}, may also form stably in highly
spin-polarized magnetic 2DEGs due to Zeeman energy considerations,
although no theoretical description of these effects exists at
present.
\begin{figure}
\epsfxsize=2.5in \center \epsfbox{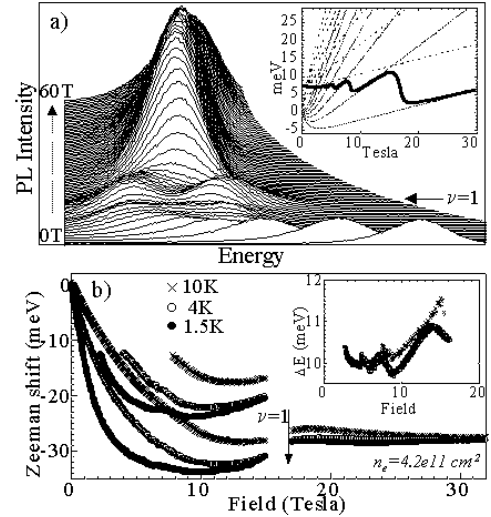} \vspace{0.1in}
 \caption{a) Characteristic evolution of the PL
spectra in high-density magnetic 2DEGs, with calculation of the
LLs and Fermi energy (inset). b) Energies of the observed PL peaks
at different temperatures, with the  ${X^-_s}$-${X^0}$ energy
splitting (inset).} \label{fig3}
\end{figure}

We turn now to results from high-density samples.  Fig. 3 shows PL
spectra and energy shifts observed in a high-density magnetic 2DEG
(${n_e=4.3\times10^{11}}$ cm${^{-2}}$, mobility=2700 cm${^2}$/Vs,
and ${g_e^{eff}(H\rightarrow0)=95}$ at 1.5K). These data are
characteristic of that obtained in samples with ${n_e}$ up to
${6.5\times10^{11}}$ cm${^{-2}}$, the highest density studied.
Again, we observe a dominant PL peak at ${H=0}$ which shifts
rapidly down in energy with applied field. However, in contrast
with the low-density 2DEGs, the high-energy satellite peak does
not appear until ~2 Tesla (at 1.5K). This satellite grows to a
peak of comparable amplitude by 12 Tesla, and exhibits similar
sensitivity to the energy and helicity of the pump laser as seen
in Fig 2; therefore we again assign these features to ${X^-_s}$
and ${X^0}$. At ${\nu=1}$ (17 Tesla), these resonances collapse
and are again replaced by a strong emission at an intermediate
energy which persists to 60T. The energy of the observed PL peaks
at 1.5K, 4K, and 10K are plotted in Fig. 3b, along with ${\Delta
E_X}$ (inset). Several features are notable. First, the ${X^0}$
peak only becomes visible at a particular {\it spin splitting} --
not field -- in support of the assertion that ${X^0}$ forms
readily only when the spin-up electron subbands depopulate to a
particular degree. In addition, the collapse of the ${X^0}$ and
${X^-_s}$ peaks occurs at ${\nu=1}$ {\it independent} of
temperature, again indicating that the drop of the Fermi energy to
the lowest LL destabilizes ${X^-_s}$. Finally, ${\Delta E_X}$
again follows the calculated Fermi energy in this sample,
exhibiting oscillations in phase with the Fermi edge.
\begin{figure}
\epsfxsize=\columnwidth \center \epsfbox{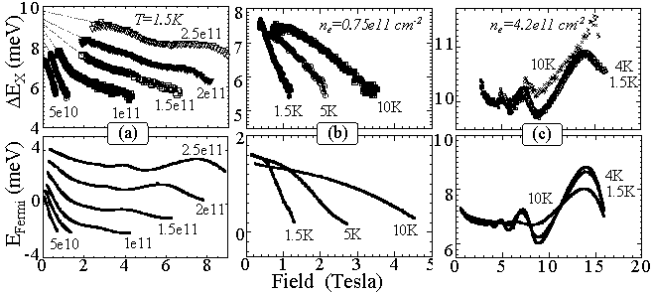}
\vspace{0.1in} \caption{Explicit dependence of the trion
ionization energy ${\Delta E_X}$ on (a) carrier density in
otherwise identical magnetic 2DEGs, and on temperature in b) low-
and c) high-density samples, all showing marked similarity to
numerical calculation of the Fermi energy.} \label{fig4}
\end{figure}

This latter behavior is unexpected but appears to be true in all
of our samples. In contrast with studies in nonmagnetic 2DEGs,
these data clearly demonstrate the relevance of both the Zeeman
energy and the Fermi energy in determining the trion ionization
energy ${\Delta E_X}$.  In Figure 4 we explicitly study this
behavior and reveal the surprising result that ${\Delta E_X}$
closely follows the energy of the Fermi surface {\it regardless}
of electron density, temperature, and applied field. Fig. 4a shows
the measured field dependence of ${\Delta E_X}$ in six magnetic
2DEGs with electron densities from ${n_e\approx5\times10^{10}}$ to
${\approx2.5\times10^{11}}$ cm${^{-2}}$. The data are plotted from
the field at which distinct ${X^0}$ and ${X^-_s}$ PL peaks first
appear, until the collapse of the PL spectra. ${\Delta E_X}$ is
seen to decrease rapidly with field at the lowest densities, but
remain roughly constant and exhibit weak oscillations at high
densities. Further, a rough extrapolation (dotted lines) reveals
that ${\Delta E_X}$ at zero field increases from ${\sim}$7meV to
10meV with carrier density. Aside from a ${\sim}$7meV difference
in overall magnitude, these features are qualitatively reproduced
by the numerical computation of the Fermi energy in these samples,
plotted in the lower graph. It is natural to associate 7 meV with
the ``bare" (${n_e\rightarrow0}$) ${X^-_s}$ binding energy, in
reasonable agreement with earlier studies in low-density,
nonmagnetic ZnSe-based 2DEGs.\cite{Astakhov} Thus, at least at
zero field, ${\Delta E_X}$ reflects the ``bare" ${X^-_s}$ binding
energy {\it plus} the Fermi energy, in agreement with a recent
viewpoint\cite{Huard} wherein the ionization process requires
removing one electron from ${X^-_s}$ to the top of the Fermi sea.

In nonzero field, the Zeeman energy reduces the ${X^-_s}$
ionization energy. The explicit temperature dependence of ${\Delta
E_X}$ in the low-density magnetic 2DEG is particularly telling
(Fig. 4b): Here, the small Fermi energy should play a minimal role
(${\varepsilon_F\sim}$1.5meV ${\ll}$ 9meV total spin splitting),
and the data should directly reveal the ${X^-_s}$ ionization
energy. At different temperatures, ${\Delta E_X}$ decreases from
its zero-field value of ${\sim}$7.5meV {\it at a rate which
depends on the Brillouin-like spin splitting}.  In this sample,
the 2DEG is almost immediately completely spin-polarized - no gas
of ``spin-up" electrons remains -- and thus the drop in ${\Delta
E_X}$ must reflect the influence of the Zeeman energy. Physically,
the energy of the spin-up electron in ${X^-_s}$ increases with
spin splitting, becoming more weakly bound, reducing ${\Delta
E_X}$ by roughly half of the total Zeeman splitting until the
${X^-_s}$ destabilizes. Within this scenario, however, the rolloff
in the slope of the data towards zero field is puzzling, possibly
indicating that the energy {\it between} the Fermi edge and the
spin-up subbands (rather than the Zeeman energy itself) may be the
relevant parameter, as the calculated Fermi energy shows precisely
the same behavior. No theoretical framework for this behavior
exists at present. Alternatively, Fig 4c shows typical data from
the high electron density sample where the Fermi energy (7.7meV)
is comparable to the total spin splitting (12.6meV). Here, the
measured ${\Delta E_X}$ clearly follows the oscillations of the
calculated Fermi energy, with no clear indication of the role
played by the Zeeman energy. We pose these questions for future
theoretical models for ${X^-_s}$ formation, which must necessarily
include the Zeeman energy and the influence of a finite Fermi
energy.

In conclusion, we have presented a systematic study of charged
exciton formation in strongly magnetic 2DEGs, wherein the giant
spin splitting dominates the cyclotron energy and the electron gas
is highly spin-polarized. The trion ionization energy ${\Delta
E_X}$ tracks the energy of the Fermi edge regardless of electron
density, temperature or applied field, highlighting the important
roles played by both the Fermi- and Zeeman energies. With
increasing electron density, the data suggest that ${\Delta E_X}$
-- at least at zero magnetic field -- reflects the ``bare"
${X^-_s}$ ionization energy of ${\sim}$7 meV {\it plus} the Fermi
energy. Studies in low density samples show that the ``bare"
${X^-_s}$ binding energy is reduced by an amount proportional to
the Zeeman energy, and in high density samples ${\Delta E_X}$
follows the oscillations of the Fermi surface as it moves between
Landau levels. Quantitative interpretation of these data must
await a more complete theory of ${X^-_s}$ formation in electron
gases. This work is supported by the NHMFL and NSF-DMR 9701072 and
9701484.

\end{document}